\documentstyle[twocolumn,aps,amssymb,epsfig]{revtex}
\begin{document}


\def\calA{{\cal A}}
\def\calB{{\cal B}}
\def\calH{{\cal H}}
\def\calO{{\cal O}}

\def\cbar{{\bar c}}
\def\nbar{{\bar n}}
\def\qbar{{\bar q}}
\def\sbar{{\bar s}}
\def\ubar{{\bar u}}


\def\etal{{\it et al.}}
\def\ibid#1#2#3{{\it ibid.} {\bf #1}, #3 (#2)}

\def\epjc#1#2#3{Eur. Phys. J. C {\bf #1}, #3 (#2)}
\def\ijmpa#1#2#3{Int. J. Mod. Phys. A {\bf #1}, #3 (#2)}
\def\jhep#1#2#3{J. High Energy Phys. {\bf #1}, #3 (#2)}
\def\mpl#1#2#3{Mod. Phys. Lett. A {\bf #1}, #3 (#2)}
\def\npb#1#2#3{Nucl. Phys. {\bf B#1}, #3 (#2)}
\def\plb#1#2#3{Phys. Lett. B {\bf #1}, #3 (#2)}
\def\prd#1#2#3{Phys. Rev. D {\bf #1}, #3 (#2)}
\def\prl#1#2#3{Phys. Rev. Lett. {\bf #1}, #3 (#2)}
\def\rep#1#2#3{Phys. Rep. {\bf #1}, #3 (#2)}
\def\zpc#1#2#3{Z. Phys. {\bf #1}, #3 (#2)}

\twocolumn[\hsize\textwidth\columnwidth\hsize\csname
@twocolumnfalse\endcsname

\title{Implications of the first observation of $B\to K_1\gamma$}
\author{Y.J. Kwon, Jong-Phil Lee\footnote{e-mail: jplee@phya.yonsei.ac.kr}}
\address{Department of Physics and IPAP, Yonsei University, Seoul, 120-749, Korea}

\tighten
\maketitle

\begin{center}
$^*$E-mail : jplee@phya.yonsei.ac.kr
\end{center}

\begin{abstract}

Implications of the recent new measurements of $B\to K_1\gamma$ by Belle are
examined.
It is shown that the new branching ratio $\calB(B\to K_1(1270)\gamma)$ requires
very large form factor compared to the theoretically predicted one.
This is an opposite case to $B\to K^*\gamma$ where theory expected larger
branching ratio.
Possible origins of the discrepancy are discussed.

\end{abstract}
\pagebreak
]


Radiative $B$ decays to kaons provide a rich laboratory to test
the standard model (SM) and probe new physics.
$B\to K^*\gamma$ is a well established process among them.
Higher resonant kaons such as $K_2^*(1430)$ are also measured by CLEO 
\cite{Coan:1999kh} and the $B$ factories \cite{Nishida:2002me,Aubert:2003zs}.
\par
Recently, Belle has announced the first measurement of $K_1(1270)$ \cite{Belle}:
\begin{equation}
\calB(B^+\to K_1^+(1270)\gamma)= (4.28\pm0.94\pm0.43)\times 10^{-5}~.
\label{Belle1270}
\end{equation}
There is also an upper bound on $K_1(1400)$ \cite{Belle}:
\begin{equation}
\calB(B^+\to K_1^+(1400)\gamma)< 1.44\times 10^{-5}~({\rm at ~90\% ~C.L.}) .
\label{Belle1400}
\end{equation}
There are many reasons to focus on the higher kaon resonances.
Firstly, they share lots of things with $B\to K^*\gamma$.
At the quark level, both of them are governed by $b\to s\gamma$;
all of the accumulated achievements of $b\to s\gamma$ can be used in radiative
$B$ decays to kaon resonances.
For example, the same operators in the operator product expansion, the same 
corresponding Wilson coefficients are available.
In addition, when the hadronic descriptions are required, the resemblance 
between $K^*$ and $K_1$ makes the analysis much easier.
Especially, the light-cone distribution amplitudes (DA) are same except the 
overall factor of $\gamma_5$ which gives rise to few differences in many 
calculations \cite{Lee:2004ju}.
\par
Secondly, $B\to K_{\rm res}(\to K\pi\pi)\gamma$ can provide a direct measurement
of the photon polarization \cite{Gronau:2001ng}.
In particular, it was shown that $B\to K_1(1400)\gamma$ can produce large 
polarization asymmetry of $\approx 33\%$ in the SM.
In the presence of anomalous right-handed couplings, the polarization can
be severely reduced in the parameter space allowed by current experimental
bounds of $B\to X_s\gamma$ \cite{Lee:2003ci}.
It was also argued that the $B$ factories can now make a lot of $B{\bar B}$
pairs enough to check the anomalous couplings through the measurement of the
photon polarization.
\par
Thirdly, theorists are now facing challenges from the discrepancy between their
predictions and experiments.
In fact, there have been noticeable theoretical advances in $B\to K^*\gamma$ 
over the last decade.
QCD corrections at next-to-leading order (NLO) of $\alpha_s$ was already
considered in \cite{Soares:1991te,Greub:1994tb,Greub:1996tg}.
Furthermore, relevant Wilson coefficients have been improved 
\cite{Adel:1993ah,Chetyrkin:1996vx} up to three-loop calculations.
Recent developments of the QCD factorization (QCDF) \cite{Beneke:2000ry} helped 
one calculate the hard spectator contributions systematically in a factorized 
form through the convolution at the heavy quark limit 
\cite{Beneke:2000wa,Beneke:2001at,Bosch:2001gv}.
$B\to K^*\gamma$ is also analyzed in the effective theories at NLO, such as
large energy effective theory \cite{Ali:2001ez} and the soft-collinear effective
theory (SCET) \cite{Chay:2003kb}.
\par
But the nonperturbative analyses should be taken into account to complete the
phenomenological explanation.
QCD sum rule or the light-cone sum rule (LCSR) is among the most reliable.
It was pointed out in \cite{Ali:2001ez}, however, that the LCSR results for the 
relevant form factor of $B\to K^*\gamma$ lead to a very large branching ratio 
compared to the measured one.
Unfortunately, there is no way to explain the gap up to now.
\par
The situation is more complicated in $B\to K_1\gamma$.
Based on the QCDF framework combined with the LCSR results, 
Ref.\ \cite{Lee:2004ju}
predicted $\calB(B^0\to K_1^0(1270)\gamma)=(0.828\pm0.335)\times 10^{-5}$ and
$\calB(B^0\to K_1^0(1400)\gamma)=(0.393\pm0.151)\times 10^{-5}$ at the NLO of
$\alpha_s$.
New measurements (Eqs.\ (\ref{Belle1270}) and (\ref{Belle1400})) certainly cast
many questions about the theoretical predictions.
Present work will be devoted to this issue.
\par
The effective Hamiltonian for $b\to s\gamma$ is
\begin{equation}
\calH_{\rm eff}(b\to s\gamma)=-\frac{G_F}{\sqrt{2}}V_{tb}V_{ts}^*
 \sum_{i=1}^{8}C_i(\mu)O_i(\mu)~,
\end{equation}
where
\begin{eqnarray}
O_2&=&(\sbar_i c_i)_{V-A}(\cbar_j b_j)_{V-A}~, \nonumber\\
O_7&=&\frac{em_b}{8\pi^2}\sbar_i\sigma^{\mu\nu}(1+\gamma_5)b_i F_{\mu\nu}~, \nonumber\\
O_8&=&\frac{g_sm_b}{8\pi^2}\sbar_i\sigma^{\mu\nu}(1+\gamma_5)T^a_{ij}b_j G^a_{\mu\nu}~,
\end{eqnarray}
are the relevant operators for present analysis.
Here $i,j$ are color indices, and we neglect the CKM element $V_{ub}V_{us}^*$ 
as well as the $s$-quark mass.
At next-to-leading order of $\alpha_s$, the decay amplitude $\calA$ is given by
\begin{equation}
\calA(B\to K_1\gamma)=-\frac{G_F}{\sqrt{2}}V_{tb}V_{ts}^*(
 C_7^{\rm eff}\langle O_7\rangle+C_2\langle O_2\rangle
 +C_8^{\rm eff}\langle O_8\rangle)~,
\end{equation}
where $\langle O_i\rangle\equiv \langle K_1\gamma|O_i|B\rangle$.
The leading contribution of $\langle O_7\rangle$ is given by
\begin{eqnarray}
\langle O_7\rangle &\equiv&
\langle K_1(p',\epsilon)\gamma(q,e)|O_7|B(p)\rangle\nonumber\\
&=&\frac{em_b}{4\pi^2}F^A_+(0)\Big[\epsilon^*\cdot q (p+p')\cdot e^*
   -\epsilon^*\cdot e^* (p^2-p^{\prime 2})\nonumber\\
&&
   +i\epsilon_{\mu\nu\alpha\beta}e^{*\mu}\epsilon^{*\nu}q^\alpha(p+p')^\beta
 \Big]~,
\end{eqnarray}
with $e^\mu$ being the photon polarization vector.
The form factor $F_+^A$ is defined by
\begin{mathletters}
\begin{eqnarray}
\lefteqn{
\langle K_1(p',\epsilon)|\sbar i\sigma_{\mu\nu}q^\nu b|B(p)\rangle}\nonumber\\
&=&
F^A_+(q^2)\Big[(\epsilon^*\cdot q)(p+p')_\mu-\epsilon^*_\mu(p^2-p^{\prime 2})\Big]
\nonumber\\
&&+F^A_-(q^2)\Big[(\epsilon^*\cdot q)q_\mu-\epsilon^*_\mu q^2\Big]\nonumber\\
&&+\frac{F^A_0(q^2)\epsilon^*\cdot q}{m_B m}\Big[
(p^2-p^{\prime 2})q_\mu-(p+p')_\mu q^2\Big]~,\\
\lefteqn{
\langle K_1(p',\epsilon)|\sbar i\sigma_{\mu\nu}\gamma_5q^\nu b|B(p)\rangle}
\nonumber\\
&=&iF^A_+(q^2)\epsilon_{\mu\nu\alpha\beta}\epsilon^{*\nu}q^\alpha(p+p')^\beta~,
\end{eqnarray}
\end{mathletters}
where $m$ and $\epsilon^\mu$ are the mass and polarization vector of $K_1$,
respectively, and $q=p-p'$ is the photon momentum.
\par

\begin{figure}
\begin{center}
\epsfig{file=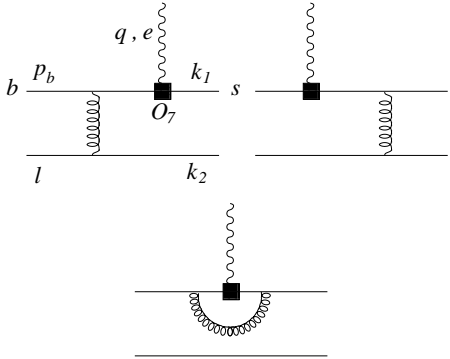,height=5cm}
\end{center}
\caption{NLO corrections to $O_7$.
These diagrams are absorbed into the weak form factor $F^A_+$.}
\label{O7NLO}
\end{figure}

\begin{figure}
\begin{center}
\epsfig{file=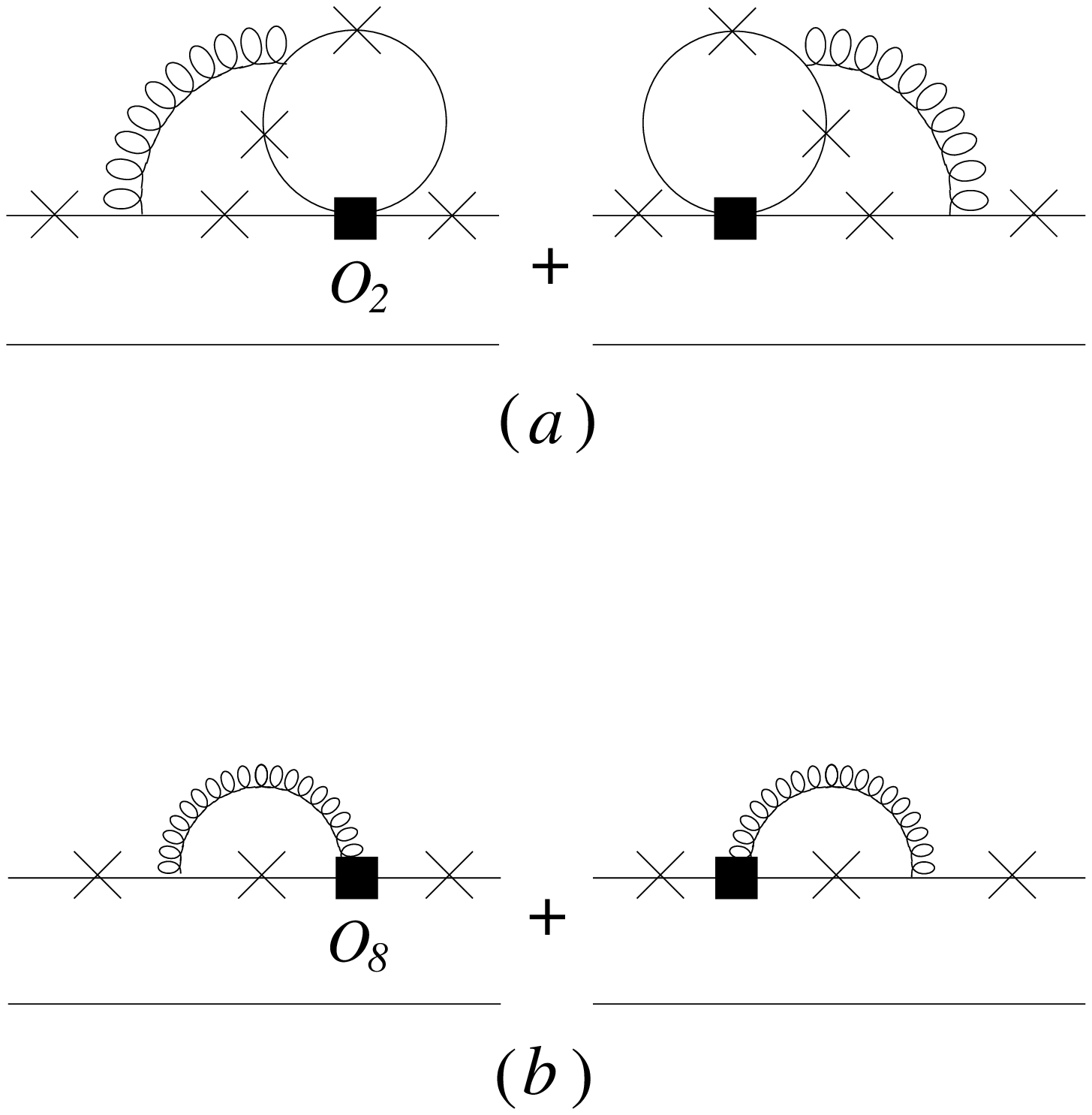,height=6cm}
\end{center}
\caption{Vertex corrections to the operators (a) $O_2$ and (b) $O_8$.
Crosses denote the possible attachment of the emitted photon.}
\label{VC}
\end{figure}

\begin{figure}
\begin{center}
\epsfig{file=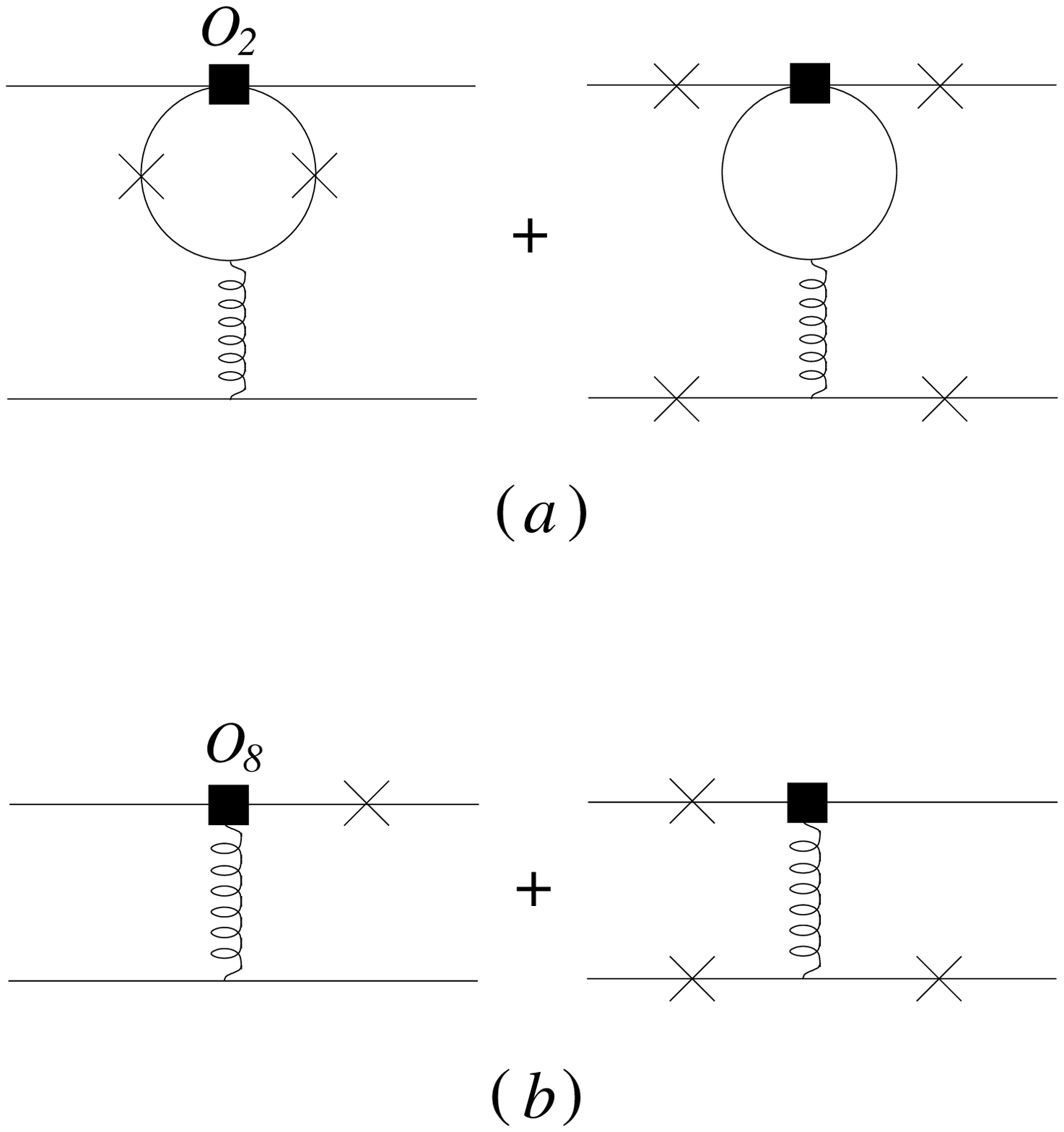,height=6cm}
\end{center}
\caption{Hard spectator interactions to (a) $O_2$ and (b) $O_8$.
First diagrams are leading contributions at the heavy quark limit.}
\label{HS}
\end{figure}

All the subleading contributions to $\langle O_7\rangle$ shown in Fig.\ 
\ref{O7NLO} are absorbed into the form factor $F_+^A$, while the corresponding
Wilson coefficient $C_7^{\rm eff}$ contains its NLO parts,
\begin{equation}
C_7^{\rm eff}(\mu)=C_7^{\rm eff(0)}(\mu)
 +\frac{\alpha_s(\mu)}{4\pi}C_7^{\rm eff(1)}(\mu)~.
\end{equation}
On the other hand, the leading order $C_2^{(0)}$ and $C_8^{\rm eff(0)}$
are sufficient for $C_2$ and $C_8$ since $O_2$ and $O_8$ contributions begin
at NLO.
The NLO contributions of $O_{2,8}$ can be written as
\begin{equation}
\langle O_i\rangle=\langle O_i\rangle_{VC}+\langle O_i\rangle_{HS}~~~
(i=2,8)~,
\end{equation}
where $\langle O_i\rangle_{VC(HS)}$ are vertex corrections (hard spectator
interactions) depicted in Figs.\ \ref{VC} (\ref{HS}).
The branching ratio of $B\to K_1\gamma$ is simply given by
\begin{eqnarray}
\lefteqn{
\calB(B\to K_1\gamma)}\nonumber\\
&=&\tau_B\frac{G_F^2\alpha m_b^2 m_B^3}{32\pi^4}\Bigg(
 1-\frac{m_A^2}{m_B^2}\Bigg)^3|F^A_+(0)|^2|V_{tb}V_{ts}^*|^2 \nonumber\\
&& \times|C_7^{\rm eff}(\mu_b)+A_{VC}+A_{HS}|^2~.
\end{eqnarray}
At the heavy quark limit,
\begin{eqnarray}
A_{VC}&=&\frac{\alpha_s(\mu_b)}{4\pi}\Bigg\{C_8^{\rm eff}(\mu_b)\Bigg[
 -\frac{32}{9}\ln\frac{m_b}{\mu_b}+\frac{4}{27}(33-2\pi^2
\nonumber\\
&& +6i\pi)\Bigg]
 +C_2(\mu_b)\Bigg[\frac{416}{81}\ln\frac{m_b}{\mu_b}+r_2\Bigg]\Bigg\}~,
\nonumber\\
A_{HS}&=&\frac{4\pi\alpha_s(\mu_H)C_F}{N_c}
 \frac{f_B f_A^\perp}{\lambda_B m_B F^A_+(0)}\Bigg\{
 C_8^{\rm eff}(\mu_H)\frac{1}{12}\langle u^{-1}\rangle_\perp
\nonumber\\
&& -C_2(\mu_H)\frac{1}{12}\left\langle
 \frac{\Delta i_5(z_0^{(c)},0,0)}{\ubar}\right\rangle_\perp\Bigg\}~.
\label{amplitudes}
\end{eqnarray}
See \cite{Lee:2004ju} for details.
\par
Keeping the hadronic parameters specifically, we have
\begin{eqnarray}
\lefteqn{
\calB(B^0\to K_1^0\gamma)}\nonumber\\
&=&0.003\times\Bigg(1-\frac{m^2}{m_B^2}\Bigg)^3\times
 \left|
 F^A_+(0)(-0.385-i0.014)\right.\nonumber\\
&& \left.+(f_A^\perp/{\rm GeV})(-0.024-i0.022)\right|^2~,
\label{master}
\end{eqnarray}
at the reference scales 
\begin{equation}
(\mu_b,\mu_H)=(m_b(m_b), \sqrt{\Lambda_H m_b(m_b)})
=(4.2~{\rm GeV}, 1.45~{\rm GeV})~.
\end{equation}
\par
It is now quite straightforward to extract the value of $F_+^A(0)$ from the new
measurements (\ref{Belle1270}) and (\ref{Belle1400}).
We have
\begin{eqnarray}
F_+^{K_1(1270)}(0)&=&0.32\pm0.03~,\nonumber\\
F_+^{K_1(1400)}(0)&<&0.19~,
\end{eqnarray}
where $f_{K_1(1270)}=0.122$ GeV, $f_{K_1(1400)}=0.091$ GeV are used
\cite{Safir:2001cd}.
These must be compared with the LCSR results \cite{Safir:2001cd}
\begin{eqnarray}
F_+^{K_1(1270)}(0)|_{\rm LCSR}&=&0.14\pm0.03~,\nonumber\\
F_+^{K_1(1400)}(0)|_{\rm LCSR}&=&0.098\pm0.02~.
\label{Safir}
\end{eqnarray}
Here we have another big difference between theory and experiment other than
$K^*$.
But the details of the differences are quite opposite.
In short,
\begin{eqnarray}
F^{K^*}_{\rm theory}>F^{K^*}_{\rm exp}~,\nonumber\\
F^{K_1}_{\rm theory}\ll F^{K_1}_{\rm exp}~.
\label{discrepancy}
\end{eqnarray}
\par
There are some candidates to explain the discrepancy.
Higher twist effects in the light-cone DA are the first one.
Usually they are process dependent, and are encoded in the coefficients of the
Gegenbauer expansion.
It is also known that they are asymptotically zero at $\mu\to\infty$ where $\mu$
is the renormalization scale.
Ref. \cite{Ali:2001ez} estimated that the non-asymptotic correction of $K^*$ at
higher twist through the Gegenbauer moments to the operator $O_8$ is 
$\sim -20\%$.
This is a bad news for $K_1(1270)$ if a similar tendency occurs for the axial
Kaons since the present analysis is based on the asymptotic form of the 
light-cone DA.
\par
The second candidate is the non-zero mass effect.
When calculating the hard spectator interactions in (\ref{amplitudes}), it is
assumed that the axial kaon is nearly massless and energetic.
Although the assumption is acceptable for $m_{K_1}\ll m_B$, the mass
hierarchy of $m_{K^*}<1~{\rm GeV}<m_{K_1}$ might impose some doubts about the
common framework for both $K^*$ and $K_1$.
Note that the chiral symmetry is broken around 1 GeV.
\par
But including non-zero mass corrections is very nontrivial.
Since the relevant large scale in $B\to K_1\gamma$ is $m_B$, possible mass
corrections will appear in the form of $m_{K_1}/m_B$.
It means that to fully appreciate the mass effects, one has to consider the
$1/m_B$ (or $1/m_b$) corrections throughout the analysis, which is not well 
established so far.
Since the discrepancy of (\ref{discrepancy}) is quite large and 
$m_{K_1}/m_B\approx 0.24$, one should expect large corrections like chiral
enhancement in non-leptonic decays at $1/m_B$.
\par
Thirdly, the framework of QCDF might not adequate for the axial koans.
The main idea of QCDF can be summarized by \cite{Bosch:2001gv}
\begin{eqnarray}
\langle V\gamma(\epsilon)|O_i|B\rangle
&=&\Bigg[F^{B\to V}(0)T^I_i \nonumber\\
&&+\int_0^1 d\xi dv~ \Phi_V(v)T^{II}_i(\xi,v)\Phi_B(\xi)\Bigg]\cdot \epsilon~,
\label{QCDF}
\end{eqnarray}
where $T^{I,II}_i$ are the hard scattering kernels.
The kernel $T^{II}_i$ is concerned with the hard spectator interactions.
The factorization of (\ref{QCDF}) holds when the hard kernels are perturbatively
calculable.
All the nonperturbative physics is encapsulated in the DAs.
A great discrepancy of (\ref{discrepancy}) suggests that this may not be the
case for $K_1$.
Some of the simple model calculations based on the heavy quark effective theory
(HQET) predict rather large branching ratios; 
see the Table V in \cite{Lee:2004ju} or refer to \cite{Ali:1992zd}.
Since the higher resonant kaons are heavy $\gtrsim 1$ GeV, it is
quite natural and attractive to consider them as heavy mesons.
In the heavy quark scheme, hard spectator interaction is inconceivable since 
almost all the momentum of initial heavy quark is transfered to the final one.
\par
We can also question the reliability of the QCD sum rule or LCSR results.
It is a common knowledge that the stability of an observable against the Borel 
parameter in the QCD sum rule gets poorer as higher resonances are involved.
Still, the problem of how to describe the higher kaon resonances remains.
It is also noticeable that the lattice calculation is very close to the QCD
sum rule result for $K^*$ \cite{Ali:2001ez,DelDebbio:1997kr}.
Much more reliable nonperturbative analyses are required in the near future.
\par
Next, possible mixing in $K_1(1270)$ and $K_1(1400)$ cannot explain the
large mismatches of (\ref{discrepancy}).
Quark model states $^3P_1$ and $^1P_1$ can mix to form physical states 
$K_1(1270)$ and $K_1(1400)$.
The form factors are now written as \cite{Cheng:2004yj}
\begin{eqnarray}
F_+^{K_1(1270)}(0)&=&Y_A(0)\sin\theta+Y_B(0)\cos\theta~,\nonumber\\
F_+^{K_1(1400)}(0)&=&Y_A(0)\cos\theta-Y_B(0)\sin\theta~,
\end{eqnarray}
where $Y_{A,B}$ are the form factors corresponding to the angular momentum
eigen states.
The enhancement from maximal mixing is only a factor of $\sqrt{2}$, assuming
$Y_A(0)\approx Y_B(0)$.
A substantial growth in $Y_{A,B}(0)$ is inevitable to explain the experimental
data.
On the other hand, the usefulness of mixing lies in the fact that it can 
naturally explain a strong suppression of $B\to K_1(1400)\gamma$.
But it is too early to say something about this point with the new upper bound
of (\ref{Belle1400}); 
the LCSR result (\ref{Safir}) is still within the boundary.
Therefore, a new observation of $B\to K_1(1400)\gamma$ is much anticipated.
\par
Finally,
it is quite unlikely that the annihilation topology would give considerable
contributions, as pointed out in \cite{Beneke:2001at,Bosch:2001gv}.
\par
In conclusion, we surveyed the implications of the first observation of 
$B\to K_1\gamma$.
The values of the relevant form factors are extracted from the experimental
data at NLO of $\alpha_s$.
We found that a very large discrepancy between theory and experiment is 
reproduced after $B\to K^* \gamma$.
Eliminating the gap will be a great challenge in theory.
Further observation by other $B$ factory as well as of $K_1(1400)$ will provide
much interest in coming days.

\begin{center}
{\large\bf Acknowledgements}\\[10mm]
\end{center}

This work was supported by the BK21 Program of the Korean Ministry of Education.


\end{document}